\documentclass[12pt]{article}
\usepackage{epsfig}               
\usepackage{graphicx}
\usepackage{amssymb}
\usepackage{cite}
\usepackage{amsmath}
\usepackage{geometry}               
\begin{document}

 \newcommand{\beq}{\begin{equation}}
\newcommand{\eeq}{\end{equation}}
\newcommand{\bea}{\begin{eqnarray}}
\newcommand{\eea}{\end{eqnarray}}
\newcommand{\beqn}{\begin{eqnarray}}
\newcommand{\eeqn}{\end{eqnarray}}
\newcommand{\beas}{\begin{eqnarray*}}
\newcommand{\eeas}{\end{eqnarray*}}
\newcommand{\defi}{\stackrel{\rm def}{=}}
\newcommand{\non}{\nonumber}
\newcommand{\bquo}{\begin{quote}}
\newcommand{\enqu}{\end{quote}}
\def\m12{\Delta m_{12}}
\def\m3{\Delta m_{13}}
\def\m24{\Delta m_{24}}


\def\de{\partial}
\def\Tr{ \hbox{\rm Tr}}
\def\const{\hbox {\rm const.}}
\def\o{\over}
\def\im{\hbox{\rm Im}}
\def\re{\hbox{\rm Re}}
\def\bra{\langle}\def\ket{\rangle}
\def\Arg{\hbox {\rm Arg}}
\def\Re{\hbox {\rm Re}}
\def\Im{\hbox {\rm Im}}
\def\diag{\hbox{\rm diag}}
\def\lsim{\mathrel{\rlap{\lower3pt\hbox{\hskip0pt$\sim$}}
    \raise1pt\hbox{$<$}}}
%
\def\gsim{\mathrel{\rlap{\lower4pt\hbox{\hskip1pt$\sim$}}
    \raise1pt\hbox{$>$}}}

\def\stroke{\vrule height8pt width0.4pt depth-0.1pt}
\def\topfleck{\vrule height8pt width0.5pt depth-5.9pt}
\def\botfleck{\vrule height2pt width0.5pt depth0.1pt}
\def\Zmath{\vcenter{\hbox{\numbers\rlap{\rlap{Z}\kern
0.8pt\topfleck}\kern
2.2pt\rlap Z\kern 6pt\botfleck\kern 1pt}}}
\def\Qmath{\vcenter{\hbox{\upright\rlap{\rlap{Q}\kern
3.8pt\stroke}\phantom{Q}}}}
\def\Nmath{\vcenter{\hbox{\upright\rlap{I}\kern 1.7pt N}}}
\def\Cmath{\vcenter{\hbox{\upright\rlap{\rlap{C}\kern
3.8pt\stroke}\phantom{C}}}}
\def\Rmath{\vcenter{\hbox{\upright\rlap{I}\kern 1.7pt R}}}
\def\Z{\ifmmode\Zmath\else$\Zmath$\fi}
\def\Q{\ifmmode\Qmath\else$\Qmath$\fi}
\def\N{\ifmmode\Nmath\else$\Nmath$\fi}
\def\C{\ifmmode\Cmath\else$\Cmath$\fi}
\def\R{\ifmmode\Rmath\else$\Rmath$\fi}


\def\QATOPD#1#2#3#4{{#3 \atopwithdelims#1#2 #4}}
\def\stackunder#1#2{\mathrel{\mathop{#2}\limits_{#1}}}
\def\stackreb#1#2{\mathrel{\mathop{#2}\limits_{#1}}}
\def\Tr{{\rm Tr}}
\def\res{{\rm res}}
\def\Bf#1{\mbox{\boldmath $#1$}}
\def\balpha{{\Bf\alpha}}
\def\bbeta{{\Bf\beta}}
\def\bgamma{{\Bf\gamma}}
\def\bnu{{\Bf\nu}}
\def\bmu{{\Bf\mu}}
\def\bphi{{\Bf\phi}}
\def\bPhi{{\Bf\Phi}}
\def\bomega{{\Bf\omega}}
\def\blambda{{\Bf\lambda}}
\def\brho{{\Bf\rho}}
\def\bsigma{{\bfit\sigma}}
\def\bxi{{\Bf\xi}}
\def\bbeta{{\Bf\eta}}
\def\d{\partial}
\def\der#1#2{\frac{\d{#1}}{\d{#2}}}
\def\Im{{\rm Im}}
\def\Re{{\rm Re}}
\def\rank{{\rm rank}}
\def\diag{{\rm diag}}
\def\2{{1\over 2}}
\def\ntwo{${\cal N}=2\;$}
\def\4N{${\cal N}=4$}
\def\none{${\cal N}=1\;$}
\def\x{\stackrel{\otimes}{,}}
\def\ba{\beq\new\begin{array}{c}}
\def\ea{\end{array}\eeq}
\def\be{\ba}
\def\ee{\ea}
\def\stackreb#1#2{\mathrel{\mathop{#2}\limits_{#1}}}

\def\Tr{{\rm Tr}}
\newcommand{\vp}{\varphi}
\newcommand{\pt}{\partial}

\setcounter{footnote}0

\vfill

\begin{titlepage}

\begin{flushright}
FTPI-MINN-06/09, UMN-TH-2437/06\\
March 16, 2006
\end{flushright}


\begin{center}

{ \Large \bf Non-Abelian  Semilocal Strings \\[2mm] in \boldmath{\ntwo}
Supersymmetric QCD  }

\end{center}
\vspace{0.1cm}
\begin{center}
 { 
 \bf    M.~Shifman$^{a}$ and \bf A.~Yung$^{a,b,c}$}
\end {center}

\begin{center}

$^a${\it  William I. Fine Theoretical Physics Institute,
University of Minnesota,
Minneapolis, MN 55455, USA}\\
$^{b}${\it Petersburg Nuclear Physics Institute, Gatchina, St. Petersburg 
188300, Russia\\
$^c${\it Institute of Theoretical and Experimental Physics, Moscow
117259, Russia}}
\end{center}

\vspace*{.3cm}
\begin{center}
{\large\bf Abstract}
\end{center}
We consider a benchmark bulk theory in four-dimensions:
 \ntwo supersymmetric QCD with the gauge group U($N$) and
$N_f$ flavors of fundamental matter hypermultiplets (quarks). 
The nature of the BPS strings in this benchmark theory crucially depends
on $N_f$. If $N_f\geq N$  and all quark masses are equal,
it supports non-Abelian BPS strings which have internal (orientational)
moduli.  If $N_f>N$ these strings become semilocal, developing additional 
moduli $\rho$ related to (unlimited) variations  of their
transverse size. 

Using the U(2) gauge group
with $N_f=3,4$ as an example, we derive an
effective low-energy theory on
the (two-dimensional) string world sheet. 
Our derivation is field-theoretic, direct and explicit: 
we first analyze the Bogomol'nyi equations for string-geometry solitons,
suggest an {\em ansatz} and solve it at large $\rho$.
Then we use this solution to  obtain the world-sheet theory.

In the semiclassical limit our result confirms
the Hanany--Tong conjecture, which rests on brane-based arguments,
 that the world-sheet theory is 
\ntwo supersymmetric U(1) gauge theory with $N$
positively and $N_e=N_f-N$ negatively charged matter multiplets and the
Fayet--Iliopoulos term determined by the four-dimensional coupling constant.
We  conclude that the  Higgs branch of this model is
not lifted by quantum effects. As a result, such strings cannot confine.

Our analysis of infrared effects,   not seen in the Hanany--Tong consideration,
shows that, in fact, the derivative expansion
can make sense only provided the theory under consideration 
is regularized in the infrared, e.g. by the quark mass differences. The world-sheet action discussed in this paper becomes a {\em bona fide} low-energy effective action
only if $\Delta m_{AB}\neq 0$.

\vspace*{.05cm}

\end{titlepage}

\section{Introduction}
\label{intro}

The recent discovery in certain supersymmetric gauge theories of 
BPS-saturated solitons 
that can be interpreted as {\sl non-Abelian strings} 
\cite{HT1,ABEKY,SYmon,HT2}
has led to a number of exciting developments 
\cite{Tong,SYnawall,J1/4,MMY,Auz,Jst,Hanany5,Sakai5,Shifman5, Auzzi5,Eto5,Tong5,Trev,Jrev}:
from confined monopoles to non-Abelian boojums,
from enhanced supersymmetry on the world sheet to
possible applications in cosmic strings and beyond.
The above-mentioned non-Abelian strings are characterized by 
non-Abelian moduli and present
a generalization of $Z_N$ strings \cite{VS,HV,Sur,SS,KB,MY,KoS}
which, in turn,
generalize the famous ANO strings \cite{ANO}.

Other topological defects with stringy geometry --- sigma model lumps ---
are known for decades. For instance, 
instantons in two-dimensional CP($N-1$) models,
lifted to four dimensions, provide probably the most clear-cut example of 
such lumps. The topological defects that interpolate between the
ANO strings and lumps are called {\sl semilocal strings}
(for a review see \cite{AchVas}). While non-trivial topology behind the 
ANO strings is related to $\pi_1 ({\rm U}(1))$, the sigma model lumps are supported by
$\pi_2({\cal T})$ where ${\cal T}$ is the target space of the sigma model at hand.
Unlike the ANO strings whose size in the transverse plane is fixed, that of the
semilocal string is a modulus. Both the ANO strings and lumps can be studied 
in a unified manner in the  framework of  gauged 
linear sigma models with a judiciously chosen 
Higgs potential.  The  special potentials which are required  here 
 are due to the Fayet--Iliopoulos terms \cite{FI}.  In an appropriate limit the 
\ntwo supersymmetric gauge theories
 with
the  Fayet--Iliopoulos term develop Higgs branches. In the low-energy limit 
effective theories on the Higgs branches become non-linear sigma models
whose target spaces have hyper-K\"ahler geometry.

In view of the recent developments it is natural to raise the question of 
{\sl non-Abelian semilocal} strings, in particular, how they emerge as BPS-saturated solitons
in \ntwo supersymmetric QCD which was previously shown to support
non-Abelian local strings. This question was first addressed in \cite{HT1}
(see also \cite{HT2})
where it was argued, on the basis of a
brane-based analysis, that the effective low-energy theory on the
world sheet of such string is given by a particular two-dimensional sigma model
with a non-compact target space presenting an example of 
 certain  special manifolds called  
{\sl toric varieties}. For an illuminating discussion see \cite{Schro}.

Needless to say, it is highly desirable to verify the
Hanany--Tong conjecture by a straightforward derivation
of the world-sheet theory for the non-Abelian  semilocal string
within the field-theoretic framework, starting from
\ntwo SQCD in the bulk. Here we carry out this derivation
in a certain limit, and demonstrate that in this limit the result 
of our direct field-theoretic calculation coincides with the Hanany--Tong formula.
{\em En route}, we clarify subtle aspects associated with
the infrared regularization of the zero modes. These aspects,
crucial for maintaining the BPS nature of the solution, were only mentioned in passing in \cite{HT1}.

We explain why, in spite of the fact that the (nonvanishing)
tension of the semilocal strings under consideration is exactly determined by the central charge of the underlying theory, the semilocal strings
do not lead to linear confinement in the conventional sense of this word.

The paper is organized as follows. In Sect.~\ref{bulkm} we present our bulk model:
\ntwo supersymmetric QCD with the gauge group U(2) and $N_f$ flavors
of fundamental  matter  fields. In Sect.~\ref{ass} we review Abelian semilocal
string solutions. Section~\ref{ssac} demonstrates that their formation leads to
deconfinement.
In Sect.~\ref{nass} we find  BPS solutions for non-Abelian
semilocal strings. Section~\ref{etsws} is devoted to  the effective theory on 
the world sheet of the non-Abelian semilocal string. In Sect.~\ref{semicre} we consider
this theory in the semiclassical limit, and in Sect.~\ref{comparison} compare this theory
to the one conjectured in \cite{HT1,HT2}. Quantum effects in the world-sheet theory are discussed in Sect.~\ref{qr}.
Section~\ref{conclu} presents our 
conclusions.

\section{The bulk model}
\label{bulkm}

The local non-Abelian strings\,\footnote{
According to the generally accepted --- albeit rather confusing --- terminology,
local as opposed to semilocal strings 
are those whose transverse size is fixed. In this sense the ANO string 
is local.}  were discovered
in \ntwo supersymmetric QCD with the gauge group SU$(N)\times$U(1)
and $N$ flavors of the matter fields. Then, if the mass terms for all matter fields are the same, the theory possesses a global flavor SU$(N)$ symmetry,
and the symmetry breaking pattern is 
$$
{\rm SU}(N)_{\rm gauge}\times{\rm SU}(N)_{\rm flavor}\to{\rm SU}(N)_{\rm diagonal}\,.
$$
To get semilocal non-Abelian strings all we  have to do is to extend the matter sector
of this theory. Namely, we must introduce $N_e$ extra flavors, so
that the total number of flavors $N_f=N+N_e$. Below we will briefly summarize the main features of our basic model, limiting ourselves to $N=2$, for simplicity.
Generalization to $N>2$ is straightforward. 

Thus, we will consider \ntwo supersymmetric QCD with the
SU(2)$\times$U(1) gauge group
and $N_f$ flavors of fundamental hypermultiplets, call them ``quarks."
If $N_f= 4$ (i.e. $N_e =2$), the $\beta$ function of the theory vanishes,
while further increase of $N_f$ leads to the loss of asymptotic freedom.
Thus, we will limit ourselves to $N_e = 1$ and 2 (i.e. $N_f= 3$ and 4).

Our theory
is perturbed by the  Fayet--Iliopoulos (FI) term of the U(1) gauge factor 
with the FI parameter $\xi$.
This parameter sets the scale of massive states in the theory, as well as
the scale of the string tension. Indeed, the $\left(\frac{1}{2}, \,\frac{1}{2}\right)$
central charge of the theory is
\begin{eqnarray}
\{Q_\alpha^I\, , \bar Q_{\dot\alpha}^J\}
&= & \delta^{IJ}\, 2\left( P_{\alpha\dot\alpha}
+ Z_{\alpha\dot\alpha}\right),\qquad I,J =1,2, \nonumber\\[0.3cm]
Z_\mu &=&\xi \int {\rm d}^3 x\, \varepsilon_{0\mu\nu\rho}
\,  \partial^\nu A^\rho\,,
\end{eqnarray}
where $A^\rho$ is the U(1) gauge field. Thus, the tension of the minimal BPS string is
$$T=2\pi\,\xi\,.$$

The field content
of SU(2)$\times$U(1) \ntwo SQCD with 
$N_f$ flavors  is as follows. The \ntwo vector multiplet 
consists of the  U(1) 
gauge fields $A_{\mu}$, SU(2)  gauge field $A^a_{\mu}$,
(here $a=1,2,3$), their Weyl fermion superpartners 
($\lambda^{1}_{\alpha}$,
 $\lambda^{2}_{\alpha}$) and
($\lambda^{1a}_{\alpha}$, $\lambda^{2a}_{\alpha}$), and 
complex  scalar fields $a$, and $a^a$, the latter in the adjoint of
SU(2). The spinorial index of $\lambda$'s runs over
  $\alpha=1,2$.  In this sector the  global SU(2)$_R$ symmetry inherent to
  the model at hand manifests itself through rotations 
$\lambda^1 \leftrightarrow
  \lambda^2$.

The quark multiplets of  the SU(2)$\times$U(1) theory consist 
of   the complex scalar fields
$q^{kA}$ and $\tilde{q}_{Ak}$ (squarks) and  
the  Weyl fermions $\psi^{kA}$ and
$\tilde{\psi}_{Ak}$,
 all in the fundamental representation of  SU(2)  gauge group.
Here $k=1,2$ is the color index
while $A$ is the flavor index, $$A=1,...,\, N_f\,,\qquad N_f = 3\,\,\,\mbox{or}\,\,\, 4.
$$
Note that the scalars $q^{kA}$ and 
${\bar{\tilde q}}^{\, kA}\equiv \overline{\tilde{q}_{Ak}}$
form a doublet under the action of the global SU(2)$_R$ group.

The bosonic part of our SU(2)$\times$U(1) theory
(in Euclidean space)  has  the form
\beqn
S&=&\int d^4x \left[\frac1{4g^2_2}
\left(F^{a}_{\mu\nu}\right)^2 +
\frac1{4g^2_1}\left(F_{\mu\nu}\right)^2
+
\frac1{g^2_2}\left|D_{\mu}a^a\right|^2 +\frac1{g^2_1}
\left|\partial_{\mu}a\right|^2 \right.
\nonumber\\[4mm]
&+&\left. \left|\nabla_{\mu}
q^{A}\right|^2 + \left|\nabla_{\mu} \bar{\tilde{q}}^{A}\right|^2
+V(q^A,\tilde{q}_A,a^a,a)\right]\,.
\label{model}
\eeqn
Here $D_{\mu}$ is the covariant derivative in the adjoint representation
of  SU(2),
while
\beq
\nabla_\mu=\partial_\mu -\frac{i}{2}\; A_{\mu}
-i A^{a}_{\mu}\, \frac{\tau^a}{2},
\label{defnabla}
\eeq
where we suppress the color  SU(2)  indices, and $\tau^a$ are the
 SU(2) Pauli matrices. The coupling constants $g_1$ and $g_2$ 
correspond to the U(1)  and  SU(2)  sectors, respectively.
With our conventions the U(1) charges of the fundamental matter fields 
are $\pm 1/2$.
The potential $V(q^A,\tilde{q}_A,a^a,a)$ in the Lagrangian (\ref{model})
is a sum of  $D$ and  $F$  terms,
\beqn
V(q^A,\tilde{q}_A,a^a,a) &=& 
 \frac{g^2_2}{2}
\left( \frac{1}{g^2_2}\,  \varepsilon^{abc} \bar a^b a^c
 +
 \bar{q}_A\,\frac{\tau^a}{2} q^A - 
\tilde{q}_A \frac{\tau^a}{2}\,\bar{\tilde{q}}^A\right)^2
\nonumber\\[3mm]
&+& \frac{g^2_1}{8}
\left(\bar{q}_A q^A - \tilde{q}_A \bar{\tilde{q}}^A -2\xi\right)^2 
\nonumber\\[3mm]
&+& \frac{g^2_2}{2}\left| \tilde{q}_A\tau^a q^A \right|^2+
\frac{g^2_1}{2}\left| \tilde{q}_A q^A  \right|^2 
\nonumber\\[3mm]
&+&\frac12\sum_{A=1}^{N_f} \left\{ \left|(a+\sqrt{2}m_A +\tau^a a^a)q^A
\right|^2\right.
\nonumber\\[3mm]
&+&\left.
\left|(a+\sqrt{2}m_A +\tau^a a^a)\bar{\tilde{q}}_A
\right|^2 \right\}\,,
\label{pot}
\eeqn
where the sum over the repeated flavor indices $A$ is implied. 
For the time being, we keep all $N_f$ mass terms $m_A$ distinct.

The first and second lines represent   $D$   terms, the third line 
the $F_a$ terms,
while the fourth and the fifth  lines represent the squark $F$ terms.
Note that the FI term does not  
break \ntwo supersymmetry \cite{HSZ,VY}.

The Fayet--Iliopoulos term triggers the spontaneous breaking
of the gauge symmetry forcing the squark fields to develop vacuum expectation
 values (VEV's). If all quark mass terms are different there are
$N_f(N_f-1)/2$ isolated vacua in which a pair of quark flavors develop
VEV's \cite{SYnawall}. We denote these vacua as $AB$-vacua, where
$A$ and $B$ are the quark flavors which develop VEV's.

Consider, say, the 12-vacuum. Up to gauge rotations the  VEV's
of the squark fields can be chosen as
\beqn
\langle q^{kA} \rangle &=&\sqrt{
\xi}\, \left(
\begin{array}{cc}
1 & 0 \\
0 & 1\\
\end{array}
\right),\,\,\,\,\,\langle \bar{\tilde{q}}^{kA}\rangle =0,
\nonumber\\[3mm]
\langle q_e^{kB}\rangle &=&\langle \bar{\tilde{q}}_e^{kB}\rangle =0,
\nonumber\\[3mm]
k&=&1,2,\qquad A=1,2\,,\qquad B=3,...,N_f\, ,
\label{qvev}
\eeqn
where we arrange the squark fields of the first two flavors
in a $2\times 2$ matrix $q$ while $q_e$ denotes extra quark flavors
(the subscript $e$ is for extra). 
The VEV's of the adjoint fields are given by
\beq
\langle a^3\rangle =-\frac{m_1-m_2}{\sqrt{2}}\,,
\,\,\,\,\langle a\rangle =-\frac{m_1+m_2}{\sqrt{2}}.
\label{avev}
\eeq 

Consider first the case $m_1=m_2$.
The color-flavor locked form of the quark VEV's in 
Eq.~(\ref{qvev}) and the absence of VEV of the adjoint scalar $a^a$ in 
Eq.~(\ref{avev}) 
results in the fact that, while the theory is fully Higgsed, a diagonal 
SU(2)$_{C+F}$ survives as a global symmetry. This symmetry involves a global
gauge transformation together with a flavor rotation  of the first two
flavors. Say, for quark fields it acts as
\beq
q\to U \,q\,U^{-1}, \qquad q_e\to U\,q_e,
\label{c+f}
\eeq
where the global gauge rotation acts from the left while the flavor rotation acts from
the right. It is clear that the vacuum (\ref{qvev}) is invariant under this transformation.
This invariance, a particular case  of the Bardakci-Halpern mechanism \cite{BarH},
 leads to the emergence \cite{ABEKY} of
orientational zero modes of the $Z_2$ strings in the model (\ref{model}).

If $N_f=3$ the  SU(2) part of the gauge group 
is asymptotically free,  implying generation of a dynamical scale
 $\Lambda$. In the infrared, if
descent to  $\Lambda$ was uninterrupted, the gauge coupling
$g_2^2$ would explode at this scale.   
Moreover,  strong coupling effects in the SU(2) subsector at the 
scale $\Lambda$ would break the  SU(2) subgroup through the
Seiberg--Witten mechanism \cite{SW1}.  Since we want to stay
at weak coupling, we assume that $\sqrt{\xi}\gg \Lambda$,  
so that the running of the SU(2) coupling  is frozen by the squark condensation
at a small value
\beq
\frac{8\pi^2}{g_2^2}=\ln{\frac{\sqrt{\xi}}{\Lambda}} +\cdots \gg 1\,.
\label{g2three}
\eeq
If $N_f=4$, the SU(2) sector of the theory is conformally invariant, and hence
the coupling $g_2$ does not run. In this case we also assume that
\beq
\frac{8\pi^2}{g_2^2}\gg 1\,.
\label{g2four}
\eeq

Now let us discuss the mass spectrum in the theory (\ref{model}). Since
both U(1) and SU(2) gauge groups are broken by squark condensation, all
gauge bosons become massive. From (\ref{model}) we get for the U(1)
gauge boson
\beq
m_{\gamma}=g_1\sqrt{\xi}
\label{phmass}
\eeq
while three gauge bosons of the SU(2) group acquire the same mass
\beq
m_{W}=g_2\sqrt{\xi}.
\label{wmass}
\eeq
It is not difficult to see from (\ref{pot}) that the adjoint fields $a$ and
$a^a$ as well as the components of the quark matrix $q$ acquire
the same masses as the corresponding gauge bosons. 
Altogether we have one long \ntwo
multiplet (eight bosonic + eight fermionic states) with the mass
(\ref{phmass}) and three long \ntwo multiplets with the mass (\ref{wmass}).
If the extra quark masses are different from $m_{1,2}$, the extra quark
flavors acquire masses determined by the mass differences
$\Delta m_{AB}=m_A-m_B$. The extra flavors become massless in the
limit $\Delta m_{AB}\to 0$, which we will consider momentarily.

\vspace{1mm}

If all quark mass terms are equal,  then the $N_f(N_f-1)/2$ isolated
vacua we had in the case of  unequal mass terms 
coalesce;  a Higgs branch develops from the common root whose location
on the Coulomb branch is given by Eq.~(\ref{avev}) with $m_1=m_2$. The 
dimension of this branch is $8\,N_e$, see \cite{APS,MY}. The Higgs branch is
 non-compact and has a hyper-K\"ahler geometry \cite{SW2,APS}.
At a generic point on the Higgs branch
the BPS-saturated string solutions do not exist \cite{RuPe};
strings become non-BPS if we move along non-compact directions \cite{EY}.
However, the Higgs branch
has a compact base manifold defined by the condition
\beq
\tilde{q}_{Ak}=0\,,\qquad A=1, ...,\,N_f\,.
\label{tildeq}
\eeq
The dimension of this manifold 
is $4\, N_e$, twice less than the total dimension of the Higgs branch.  The real dimension of the
base manifold is 4 for $N_f=3$ and 8 for $N_f=4$.
The BPS-saturated string solutions exist on the base manifold of the Higgs
branch, therefore, the vacua  belonging 
to the base manifold are our prime focus.

\vspace{1mm}

The base of the Higgs branch can be generated by flavor rotations of the 
12-vacuum (\ref{qvev}). For $N_f=3$, the flavor rotations generate the  manifold
\beq
\frac{{\rm SU}(3)}{{\rm SU}(2)_{C+F} \times {\rm U}(1)}
\eeq
where SU(2)$_{C+F}$ is a global unbroken color-flavor rotation which involves
the first two flavors, while the U(1) factor stands for the unbroken 
U(1) flavor rotation
of the third flavor. Dimension of this  quotient is four, indeed.
For $N_f=4$, the base of the Higgs branch is isomorphic to
\beq
\frac{{\rm SU}(4)}{{\rm SU}(2)_{C+F}\times {\rm U}(2)}
\eeq
where the U(2) factor stands for  flavor rotations of the third and 
fourth flavors
left unbroken by (\ref{qvev}). Dimension of this quotient is 8, as was expected.

\section{Abelian semilocal strings}
\label{ass}

The flux tube (string) solutions on the Higgs branches (which are typical
for multi-flavor theories) usually are not conventional ANO strings, but, rather,
semilocal strings (see \cite{AchVas} for a review).
Here we give a brief introduction to semilocal strings in a simplified
non-supersymmetric environment in the U(1) model.

As was mentioned, the semilocal string
interpolates between the ANO string and two-dimensional  sigma-model instanton
lifted to four dimensions (this is referred to as
lump). The semilocal string possesses
and additional zero mode associated with string's transverse size $\rho$.
 At  $\rho\to 0$ we have the ANO string while
at $\rho \to\infty$ it becomes a lump.
At non-zero $\rho \neq 0$ the profile functions of the semilocal
string fall-off at infinity as inverse powers
of the distance,
instead of the exponential fall-off characteristic to ANO strings at $\rho=0$.
This leads to a dramatic physical effect ---  
semilocal strings, in contradistinction to the ANO strings,
do not support linear confinement (see below).

The simplest model where the semilocal strings appear is the Abelian Higgs
model with {\sl two} complex flavors 
\begin{equation}
S_{\rm AH}=\int d^4x\left\{\frac1{4g^2}\,F^2_{\mu\nu}+|\nabla_\mu
q^A|^2+\frac{g^2}{8}\left( |q^A|^2-\xi\right)^2\right\}\,.
\label{ah2fl}
\end{equation}
Here $A=1,2$ is the flavor index. The model contains only bosonic fields,
it is not supersymmetric. The scalar potential in Eq.~(\ref{ah2fl})
is inspired by supersymmetric models with the Fayet--Iliopoulos term \cite{FI}.
The covariant derivative is defined as
$$\nabla_\mu=\partial_\mu -\frac{i}{2}\; A_{\mu}\,,$$
so that the electric charge of both quarks is $1/2$. 

\vspace{1mm}

If $\xi > 0$ the scalar fields develop VEV's breaking the U(1) gauge group.
The photon field gets Higgsed, and gets a mass, together with one real scalar.
For the particular choice of the quartic coupling presented in Eq.~(\ref{ah2fl})
this scalar has the same mass as the photon, since our toy model (\ref{ah2fl})
is a bosonic reduction of  an $\,{\cal N}=1$ supersymmetric theory in which
the vortices are BPS-saturated. Two other scalars remain massless.
 
 \vspace{1mm}
 
The topological reason for the existence of the ANO vortices is that
 $\pi_1[{\rm U}(1)]=Z$. On the other hand  we can go to
the low-energy limit in (\ref{ah2fl})  assuming that $m_{\gamma}\to\infty$ and
integrating out the massive photon and the real massive scalar field.
This will lead us to a four-dimensional sigma model on 
the vacuum manifold $$|q^1|^2 + |q^2|^2 =\xi\,.$$ This vacuum manifold has
dimension $4-1-1=2$, where we subtract
one real condition and one gauge phase. (One can always choose the gauge 
in which,
say, $q^1$ is real). The target space of the sigma model 
represents two-dimensional sphere $S_2$.  Thus, the low-energy limit
of the theory (\ref{ah2fl}) is the O(3) sigma model.
Now recall that
$$\pi_2[S_2]=\pi_1[{\rm U}(1)]=Z\,.$$
This is the topological reason for
the existence of instantons in  two-dimensional
O(3) sigma model.  Lifted to four dimensions they become
string-like objects (lumps).

So, now the question is: what is the relation between the ANO flux tubes
of scalar QED (\ref{ah2fl}) and the lumps of the O(3) sigma model?
It is clear that the model (\ref{ah2fl}) supports the ANO strings. Say, if we put the
the second flavor field 
$q^2 =0$, this model reduces to the   standard framework
for the critical ANO strings.

However, it turns out (see \cite{H}) that the ANO solution in the model at
hand
has a zero mode associated with exciting the second flavor.
This zero mode is parametrized by a 
complex parameter $\rho$ where $|\rho|$  plays the
role of the transverse size of the string while the phase of $\rho$
describes a U(1) rotation angle in O(3). To see that this zero mode indeed occurs
let us examine the solution.
To this end we will modify the standard parametrization
\cite{ANO}  for the ANO string,  including the second flavor,
\begin{eqnarray}
\label{profsl}
q^1(x) &=& \phi (r)\,  e^{i\,\alpha}\ ,\nonumber\\[2mm]
q^2(x) &=& \chi (r)\, ,\nonumber\\[2mm]
A_i(x) &=&\! \!-2\epsilon_{ij}\,\frac{x_j}{r^2}\ [1-f(r)]\, ,\quad i,j=1,2\,,
\end{eqnarray}
where $r$ and $\alpha$ are polar coordinates in the perpendicular (1,2)-plane.
We assume that the string is aligned along the $x_3$ axis.
Note, that the second flavor does {\sl not} wind at infinity. Therefore,
its boundary condition at infinity is $\chi (\infty)=0$, while at $r=0$ the function
$\chi$ need not vanish.
 The boundary conditions for other profile functions are
\begin{eqnarray}
&& \phi (0)=0\ ,
\qquad ~f(0)=1\ , \nonumber\\ [3mm]
&& \phi (\infty)=\sqrt{\xi}\ ,
  \quad
f(\infty)=0\,.
\label{bc}
\end{eqnarray}
These boundary conditions ensure that  $|q^A|^2\to {\xi}$ at 
infinity, while  the string carries one unit of the magnetic flux and has a finite tension.
The first order Bogomol'nyi equations \cite{B} for the profile functions take
the form
\beqn
&& r\frac{\rm d}{{\rm d}r}\,\phi (r)- f(r)\,\phi (r)\ =\ 0\ ,
\nonumber\\[3mm]
&& r\frac{\rm d}{{\rm d}r}\,\chi (r)- (f(r)-1)\,\chi (r)\ =\ 0\ ,
\nonumber\\[3mm]
 && 
 -\frac1r\,\frac{\rm d}{{\rm d}r} f(r)+\frac{g^2}{4}\,
\left\{ \phi^2(r)+\chi^2(r)-\xi\right\}  =\ 0\ .
\label{foesl}
\eeqn
The
ANO string  solution implies that $\chi=0$. In fact, the
second equation in (\ref{foesl})  can  be solved in the general form,
\beq
\chi=\frac{\rho}{r}\,\phi\,,
\label{chi}
\eeq
expressing $\chi$ in terms of $\phi$ and an arbitrary complex parameter $\rho$.
If we set $\rho =0$, the second flavor profile function indeed vanishes.
However, at $\rho \neq 0$ it does not.

The solution to  equations (\ref{foesl})  at   $\rho\neq 0$
is very different \cite{H,LeSa} from that for the ANO string. It has a long range power
fall-off at infinity for all profile functions. In particular,
in the limit of a very
large transverse size of the string,  $\rho\gg 1/g\sqrt{\xi}$, the
solution has the form
\beqn
\phi(r)&=&\sqrt{\xi}\frac{r}{\sqrt{r^2+|\rho|^2}}\,,
\nonumber\\[3mm]
\chi(r)&=&\sqrt{\xi}\frac{\rho}{\sqrt{r^2+|\rho|^2}}\,,
\nonumber\\[3mm]
f&=&\frac{|\rho|^2}{r^2+|\rho|^2}\,.
\label{lump}
\eeqn
This solution certainly has the same tension as the ANO string,
\beq
\label{lumpten}
T=2\pi\xi\,.
\eeq
Equation (\ref{lump}) is valid at distances $r\gg
1/g\sqrt{\xi}$.
Examining Eq.~(\ref{lump}) we
see that the scalar fields in this solution lie on the vacuum manifold
$|q^A|^2=\xi$ at any $r$ as long as these expressions are valid.
That is not the case for the ANO string.
Inside the ANO string the scalar fields tend to zero; they approach
the vacuum point only at $r\to\infty$.

The fact that $|q^A|^2=\xi$ at any $r$  means that we can relate the solution
(\ref{lump}) to the O(3) sigma model lump. To this end we  use
the  standard relation between the O(3) and CP(1) model variables,
\beqn
&&
\frac1{\xi}\,\bar{q}_A(\tau_3)^A_B \, q^B =\frac{1-|w|^2}{1+|w|^2}\,,
\nonumber\\[3mm]
&&
\frac1{\xi}\,\bar{q}_A(\tau_1)^A_B \, q^B =2\,\frac{{\rm Re}\,w}{1+|w|^2}\,,
\nonumber\\[3mm]
&&
\frac1{\xi}\,\bar{q}_A(\tau_2)^A_B \, q^B =2\, \frac{{\rm Im}\,w}{1+|w|^2}\,,
\label{o3rel}
\eeqn
where $\tau_{1,2,3}$ are flavor Pauli matrices.
With  this substitution, the low-energy limit of the action (\ref{ah2fl})
reduces to that of the following O(3) sigma model:
\beq
\label{o34d}
S_{\,\rm eff}=\xi\,\int d^4 x \, \frac{|\partial_{\mu}w|^2}{(1+|w|^2)^2}\, ,
\eeq
with $\xi$ playing the role of the coupling constant.
In this model the standard lump solution 
centered at the origin takes the form
\beq
\label{inst}
w_{\,\rm lump}=\frac{\rho}{x_1+ix_2}\,,
\eeq
where the complex modulus $\rho$ is associated with lump's size.
Re-expressing this solution in terms of the quark fields
through (\ref{o3rel}) we recover
the solution (\ref{lump}). This is a direct 
and transparent demonstration of the fact 
that the semilocal string
in the limit of large $\rho$  is described by the lump
solution of the O(3) sigma model.

\section{Semilocal strings and confinement}
\label{ssac}

The semilocal strings discussed in Sect.~\ref{ass} are BPS-saturated.
As was mentioned, their tension $T=2\pi\xi$ irrespective of the value of $\rho$.
At first sight it might seem that they must support  linear confinement of  monopoles,
much in the same way as the ANO strings.
The transverse size of the ANO string is  $\sim 1/g\sqrt{\xi}$;  if the string length $L\gg g\sqrt{\xi}$  the energy of this configuration is 
\beq
V(L)=T\, L\,.
\label{conf}
\eeq
 This linear potential ensures confinement of monopoles. Needless to say, if
$L\ll g\sqrt{\xi}$, there is no linear potential. 

For semilocal strings the transverse size is a modulus.
However, the adequate formulation of the problem is as follows.
Assume we have a monopole-antimonopole pair separated by a distance $L$.
Then the string to which the (anti)monopoles are attached has length $L$.
If $L$ is finite, the collective coordinate $\rho$ looses its moduli
status. At small $\rho$ a
slightly  negative mode develops, 
since it is energetically favorable
to increase $\rho$. This instability in $\rho$ will be regulated by the 
string length
parameter $L$ itself. In other words, the transverse size of the finite-length
semilocal string will be stabilized at $\rho\sim L$.

Clearly, the problem becomes three-dimensional.  The
monopole flux is not trap\-ped now inside  a narrow flux tube. Instead, it is
freely spread over a large three-dimensional volume  of size  
$\,\sim L^3$. This produces a Coulomb-type potential between 
the probe monopole and
antimonopole
\beq
V(L)\sim 1/L\, ,
\label{coulomb}
\eeq
up to possible logarithms.
The energy of this configuration is lower than the one of the stringy
configuration (\ref{conf});
therefore, it is energetically favored. The semilocal string increases its size 
with $L$ and effectively disintegrates,
giving place to a Coulomb-type  interaction.
It should be added that lattice studies
confirm \cite{Leese} that the semilocal string thickness tends to increase upon small
perturbations.
 
Thus, the formation of semilocal strings  on  the Higgs
branches, replacing the ANO strings existing when we deal with isolated vacua,
leads to a dramatic physical effect --- deconfinement. 
Below we turn to non-Abelian semilocal strings in the theory (\ref{model})
with $N_f=3,4$ to find out whether or not  the size modulus $\rho$  is
lifted in quantum theory  after taking into account a strong coupling 
of the size zero mode to  
interacting orientational zero modes.

\section{Non-Abelian semilocal strings}
\label{nass}

If the above material can be viewed, in a sense, as an extended introduction,
now we turn to construction and analysis of the semilocal non-Abelian
strings in earnest.

Local non-Abelian BPS-saturated strings were  found in \ntwo QCD with
the  gauge group
SU($N)\times$U(1)  in\cite{HT1,ABEKY,SYmon,HT2}. As was mentioned,
their crucial
feature is the occurrence of orientational zero modes associated with rotation
of the magnetic flux inside the SU($N$) group. 
The key ingredient in the construction of non-Abelian strings
is the presence of an unbroken 
global non-Abelian color-flavor subgroup (SU(2)$_{C+F}$
in the $N=2$ case   considered here), see
Eq.~(\ref{c+f}). This symmetry  is broken, however, on the string solution.
The Goldstone modes associated with the above breaking become orientational
zero modes of the non-Abelian string.

For isolated vacua in \ntwo QCD with the gauge group SU(N)$\times$U(1) 
and $N_f=N$ the non-Abelian string solutions were explicitly
found in \cite{ABEKY}. It was done in two steps. First, a  $Z_N$
string solution was obtained. Then, rotations from SU(N)$_{C+F}$ were applied to 
this solution, producing a family of solutions parametrized by
the orientational moduli. Now, we will generalize the procedure of \cite{ABEKY}
to cover the case of the semilocal strings in the theory (\ref{model}).
We will consider $N=2$ and $N_e=1$ and 2.

We will focus on string solutions on the base of the Higgs branch defined
by the condition (\ref{tildeq}); hence, we assume that $\tilde{q}=0$
on the string solution. For the ANO strings the adjoint fields $a$ and $a^a$ played
no role in the solution provided that $m_1=m_2$. Assuming that the same is true for 
the semilocal strings we can simplify our theory by dropping 
the adjoint field (in addition to $\tilde{q}$) from the action (\ref{model}).

Then the Bogomol'nyi completion \cite{B} of the action leads to  the
following first-order equations:
\begin{eqnarray}
& &  F^{*a}_{3}+
     \frac{g^2_2}{2}\, 
\left(\bar{q}_A\tau^a q^A\right)=0   , \qquad  a=1,2,3;
\nonumber \\[3mm]
& &   F^{*}_{3}+
     \frac{g^2_1}{2}\, \
\left(|q^A|^2-2\xi \right)=0;
\nonumber \\[4mm]
 & &   ( \nabla_1  +i \, 
\nabla_2)\, q^A=0, \qquad
\label{F38}
\end{eqnarray}
where 
\beq
F^*_m =\frac{1}{2}\, \varepsilon_{mnk}\, F_{nk}\,,\qquad m,n,k = 1,2,3\,.
\label{fstar}
\eeq
(see Ref.~\cite{ABEKY}).

The minimal or elementary $Z_2$ string emerges
when  the first flavor has the unit winding number while
the second flavor does not wind at all, to be referred to as the
(1,0) string. Extra flavors have vanishing VEV's and 
cannot wind, see Eq.~(\ref{qvev}). Needless to say,
there is   another $Z_2$ string solution in which the second flavor has the unit winding number while
the first flavor does not wind. This is called the (0,1) string. 
Together, they form a set of two $Z_2$ strings.

The conventional Abelian string forces both flavors to have the unit winding number;
therefore,  it must be viewed as the
(1,1) string \cite{ABEKY,SYmon}. Its magnetic flux and tension are twice larger 
than those of the $Z_2$ strings.
Consider for definiteness the (1,0)
string.  To find appropriate  solution to 
first-order equations (\ref{F38}) we modify the 
{\em ansatz} (\ref{profsl}) as follows (see also \cite{ABEKY}):
\beqn
q(x)
&=&
\left(
\begin{array}{cc}
  e^{ i \, \alpha  }\phi_1(r) & 0  \\
  0 &  \phi_2(r) \\
  \end{array}\right),
\nonumber\\[4mm]
q_e (x)
&=&
\left(
\begin{array}{cc}
  \chi_1(r)& 0   \\
  0 & \chi_2(r) \\
  \end{array}\right),
\nonumber\\[4mm]
A^3_{i}(x)
&=&
 -\varepsilon_{ij}\,\frac{x_j}{r^2}\
\left(1-f_3(r)\right),\qquad i,j=1,2\,,
\nonumber\\[4mm]
A_{i}(x)
&=&
 - \varepsilon_{ij}\,\frac{x_j}{r^2}\
\left(1-f(r)\right)\,,
\label{sol}
\eeqn
where the
real  profile functions $\phi_1$, $\phi_2$ 
and complex functions $\chi_1$, $\chi_2$ for the scalar fields, as well as
$f_3$, $f$ for the gauge fields, depend only on $r$.
The above {\em  ansatz} refers to
$N_f=4$. The case $N_f=3$ can be readily obtained from (\ref{sol})
by truncating  the $2\times 2$ matrix for $q_e$, replacing it by  a two-component
column  with the  entries $\left(\begin{array}{cc}\chi_1 \\ 0\end{array}\right)$.

\vspace{2mm}

Applying this {\em ansatz} one can rearrange the first-order
equations (\ref{F38}) in  the form
\beqn
&&
r\frac{d}{{d}r}\,\phi_1 (r)- \frac12\left( f(r)
+  f_3(r) \right)\phi_1 (r) = 0\, ,
\nonumber\\[4mm]
&&
r\frac{d}{{ d}r}\,\phi_2 (r)- \frac12\left(f(r)
-  f_3(r)\right)\phi_2 (r) = 0\, ,
\nonumber\\[4mm]
&&
r\frac{d}{{d}r}\,\chi_1 (r)- \frac12\left( f(r)
+  f_3(r) -2\right)\chi_1 (r) = 0\, ,
\nonumber\\[4mm]
&&
r\frac{d}{{d}r}\,\chi_2 (r)- \frac12\left( f(r)
-  f_3(r) \right)\chi_2 (r) = 0\, ,
\nonumber\\[4mm]
&&
-\frac1r\,\frac{ d}{{ d}r} f(r)+\frac{g^2_1}{2}\,
\left[\left(\phi_1(r)\right)^2 +\left(\phi_2(r)\right)^2
+\left|\chi_1(r)\right|^2+ \left|\chi_2(r)\right|^2-2\xi \right] =
0\, ,
\nonumber\\[4mm]
&&
-\frac1r\,\frac{d}{{ d}r} f_3(r)+\frac{g^2_2}{2}\,
\left[\left(\phi_1(r)\right)^2 -\left(\phi_2(r)\right)^2
+\left|\chi_1(r)\right|^2 -\left|\chi_2(r)\right|^2\right]  = 0
\, .
\label{foe}
\eeqn
Again, we hasten to add that these equations are written down for $N_f=4$.
If $N_f=3$ one should put  $\chi_2=0$ in Eq.~(\ref{foe}).

Next, we need to specify the boundary conditions
which would determine the profile functions in these equations. It is not difficult to see
that one must require
\beqn
&&
f_3(0) = 1\, ,\qquad f(0)=1\, ;
\nonumber\\[4mm]
&&
f_3(\infty)=0\, , \qquad   f(\infty) = 0
\label{fbc}
\eeqn
for the gauge fields, while the boundary conditions for  the
squark fields are
\beqn
\phi_1 (\infty)
&=&
\sqrt{\xi}\,,\qquad   \phi_2 (\infty)=\sqrt{\xi}\,,
\qquad \phi_1 (0)=0\, 
\nonumber\\
\chi_1(\infty)
&=&
0\,, \qquad \chi_2(\infty)=0\,.
\label{phibc}
\eeqn
Note that since the field $ \phi_2 $ does not wind, it need not vanish
at the
origin, and it does not.

As in the Abelian case, the equations for $\chi$'s can be solved in the
general form,
\beq
\chi_1\sim\frac{1}{r}\,\phi_1\,,\qquad\chi_2\sim\frac{1}{r}\,\phi_2\,.
\label{chi12}
\eeq
Now let us consider the solutions to the first-order equations assuming
the size $\rho$ of the semilocal string to be very large, 
\beq
|\rho|\gg\frac1{m_{\gamma}}\,,\,\,\frac1{m_{W}},
\label{largerho}
\eeq
where the masses of the gauge bosons are given in Eqs.~(\ref{phmass}) and (\ref{wmass}) and we assume that $m_{\gamma}\sim m_{W}$.
For the case $N_f=3$ we have
\beqn
q (x)
&=&
\left(
\begin{array}{cc}
  e^{ i \, \alpha  }\phi(r) & 0  \\
  0 &  \sqrt{\xi} \\
  \end{array}\right),
\nonumber\\[4mm]
q_e (x)
&=&
\left(
\begin{array}{c}
  \rho  \\
  0  \\
  \end{array}
\right)\frac{\phi(r)}{r},
\nonumber\\[4mm]
f_3
&=&
f,
\label{3z2}
\eeqn
where the profile functions $\phi (r)$ and $f(r)$ are presented in Eq.~(\ref{lump}).
The modulus  $\rho$ is the complexified size of the lump. We see that the second flavor
essentially plays no role in the solution since it is equal to its VEV everywhere,
$|q^2|^2\equiv\xi$.
The first and the third flavor vary along the base of the Higgs branch,
\beq
|q^1|^2+|q^2|^2+|q^3|^2 =2\xi\quad\mbox{at all}\quad r\, .
\label{basehiggs}
\eeq

As in the Abelian case, this solution is, in fact,  a lump of the low-energy 
four-dimensional sigma model on the base of the Higgs branch. Clearly, we can
interchange the first and the second flavors, simultaneously flipping the sign
of $f_3$, to get the other $Z_2$ string solution, namely, the (0,1) string.

If we have four flavors the solution takes the form
\beqn
q (x)
&=&
\left(
\begin{array}{cc}
  e^{ i \, \alpha  }\phi(r) & 0  \\
  0 &  \sqrt{\xi} \\
  \end{array}\right),
\nonumber\\[4mm]
q_e(x)
&=&
\left(
\begin{array}{cc}
  \rho_1 & \rho_2  \\
  0 & 0 \\
  \end{array}
\right)\frac{\phi(r)}{r},
\nonumber\\[4mm]
f_3
&=& f\,,
\label{4z2}
\eeqn
where we use the possibility of arbitrary U(2) flavor rotations for the third and
fourth flavors,
which ensures the parametrization of the solution by two complex numbers
$\rho_1$ and $\rho_2$. In this case, the the size of the string $\rho$ entering
the profile functions (\ref{lump}) is given by
\beq
|\rho|^2 \equiv |\rho_1|^2 + |\rho_2|^2 \,.
\label{size}
\eeq

\vspace{2mm}

The elementary $Z_2$ strings give rise to the
 non-Abelian strings  provided the condition
$m_1=m_2$ is satisfied \cite{HT1,ABEKY,SYmon,HT2}.
Orientational 
moduli are generated corresponding to spontaneous
breaking of the ``flat"
vacuum SU($2$)$_{C+F}$ symmetry on 
the solutions (\ref{3z2}) and (\ref{4z2}).
The color-flavor locked SU($2$)$_{C+F}$ is broken
down  to U(1). This implies 
 two  orientational moduli  ($2(N-1)$ for the bulk theory with the SU(N)$\times$U(1) gauge group).

To obtain the semilocal non-Abelian string solution from the $Z_2$ string 
(\ref{3z2}) and (\ref{4z2})  we apply the diagonal color-flavor 
rotation (\ref{c+f}) preserving
the vacuum (\ref{qvev}). To this end
it is convenient to pass to the singular gauge where the scalar fields have
no winding at infinity, they are aligned, while the string magnetic  flux
is saturated near   
the origin. In this gauge we have for the gauge fields
\begin{eqnarray}
A^a_{i}(x)
&=&
S^a \,\varepsilon_{ij}\, \frac{x_j}{r^2}\,
f_3(r)\, ,
\nonumber \\[4mm]
A_{i}(x)
&=&
 \varepsilon_{ij} \, \frac{x_j}{r^2}\,
f(r)\, ,
\label{strg}
\end{eqnarray}
where $ S^a$ is a moduli vector
defined as
\beq
S^a\tau^a=U \tau^3 U^{-1},\;\;a=1,2,3,\qquad \sum_{a=1}^3\, S^a\,S^a =1.
\label{s}
\eeq
and $U$ is a matrix from SU($2$)$_{C+F}$.
For $N_f=3$, the quark fields have the form
\beqn
q(x)
&=&
U\,\left(
\begin{array}{cc}
  \phi(r) & 0  \\
  0 &  \sqrt{\xi} \\
  \end{array}\right)\,U^{-1}\, ,
\nonumber\\[4mm]
q_e(x)
&=&
U\,\left(
\begin{array}{c}
  \rho  \\
  0  \\
  \end{array}
\right)\frac{\phi(r)}{r},
\label{3strq}
\eeqn
while for $N_f=4$ we get
\beqn
q(x)
&=&
U\,\left(
\begin{array}{cc}
  \phi(r) & 0  \\
  0 &  \sqrt{\xi} \\
  \end{array}\right)\,U^{-1}\, ,
\nonumber\\[4mm]
q_e(x)
&=&
U\,\left(
\begin{array}{cc}
  \rho_1 & \rho_2  \\
  0 & 0 \\
  \end{array}
\right)\frac{\phi(r)}{r}\,.
\label{4strq}
\eeqn

\section{Effective theory on the string world sheet}
\label{etsws}

The fact that a proper {\em ansatz} for non-Abelian semilocal strings 
can be found
and solved (in an explicit analytic form at $|\rho |\gg m_{\gamma ,\, W}^{-1}$)
is tantalizing by itself. This is not the end of the story, however.
Our next task is to derive the world-sheet theory of moduli.
In this section we will address this issue. To this end, we will
promote the string moduli  parameters to 2D
fields on the string world sheet, assuming adiabatic dependence on the 
world-sheet coordinates. As usual, the translational moduli decouple;
we will ignore them hereafter. We will focus
on internal dynamics of the string at hand. For local non-Abelian strings 
occurring in the
isolated vacua (such strings are obtained if $N_f=2$ in the action (\ref{model}))
the internal moduli are given by the orientation vector $S^a$. 
The low-energy 
world-sheet theory governing these orientational moduli is  CP(1), with 
the action\,\footnote{In this and many subsequent expressions
for the world-sheet action we omit the fermion part.}
\beq
S^{(1+1)}_{N_f=2}=\frac{\beta}{2}\,\, \int dt\ dz \,\,(\pt_k S^a)^2\,,
\label{o3}
\eeq
where $k=0,3$, while the two-dimensional coupling constant $\beta$ is related
to the four-dimensional coupling as
\beq
\beta =\frac{2\pi}{g^2_2}\,,
\label{beta}
\eeq
at the scale $\sqrt{\xi}$ which determines the string transverse size
\cite{ABEKY,SYmon}.

Now, we introduce one or two extra flavors,
$N_f=3,4$, which triggers the conversion
of the  non-Abelian local string into  semilocal. In addition to the
orientational moduli $S^a$, the semilocal string acquires the size moduli $\rho_i$, see
Eqs.~(\ref{strg}), (\ref{3strq}) and (\ref{4strq}). Below we 
study interplay between the orientational and size moduli and 
derive an effective world-sheet theory ---
first, for the case of equal quark masses $\Delta m_{AB}=0$ and, later,
introducing small mass differences. The latter are crucial for infrared regularization.

\subsection{The equal mass case}

Assume  that the orientational and size collective coordinates $S^a$
and $\rho_i$
are slow-varying functions of the string world-sheet coordinates
$x_k$ where  $k=0,3$. Then these moduli  become fields of a
(1+1)-dimensional sigma model on the string world sheet. Since
they parametrize the impact of the
string zero modes, no potential term emerges. 
We must derive only  the kinetic term.

To obtain the   kinetic term  we substitute our solution, which depends
on the moduli $ S^a$ and  $\rho_i$
in the action (\ref{model}) assuming  that
the moduli fields acquire a dependence on the coordinates 
$x_k$ via $S^a(x_k)$ and
$\rho_i(x_k)$. As in the case of local non-Abelian strings
\cite{ABEKY,SYmon}, we will have to modify our string solution 
extending our {\em ansatz} to include
the $k=0,3$ components of the SU(2) gauge field,
\beq
A_k=-\, \varepsilon^{abc}\, S^b \, \pt_k S^c \,\omega (r)\, , \qquad k=0, 3\,,
\label{Ak}
\eeq
where a new profile function $\omega (r)$ is introduced. 

The function $\omega (r)$ in Eq.~(\ref{Ak}) is
determined  through a minimization procedure 
which generates $\omega$'s own equation of motion.
Note that
 $\omega $ must satisfy the  boundary conditions
\beq
\omega (\infty)=0\,, \qquad \omega (0)=1\,,
\label{bcomega}
\eeq
which ensure  finiteness of the contribution  to the action due to the
gauge kinetic term Tr$\,F_{ki}^2$. 

\vspace{2mm}

Let us start from $N_f=3$. Symmetry arguments forbid 
mixed kinetic terms involving both, derivatives of $S^a$ and derivatives of
$\rho$.  
Hence, we can proceed in  two steps. First, assume that $S^a$ has an adiabatic dependence on the
world sheet coordinates while $\rho$ is constant. This will give us 
a part of two-dimensional action with the kinetic term for $S^a$. Then,
we  will assume, instead,  that only the field $\rho$ is $x_k$-dependent,
ignoring the  $x_k$ dependence of  $S^a$. This will give rise to the 
kinetic term for $\rho$.

Substituting the string solution (\ref{strg}), (\ref{3strq}) and (\ref{Ak})
in the action (\ref{model}) and ignoring the $x_k$ dependence of $\rho$
 we get the CP(1) model
(\ref{o3}) with the coupling constant $\beta=\beta_S$
where now $\beta_S$ is given by the following normalizing integral:
\beq
\beta_S = 
\frac{2\pi}{g_2^2}\,m_W  \int_0^{\infty}
rdr\left\{ \omega^2
+\left(1-\omega \right)\left(\frac{\phi}{\xi}-1\right)^2
+ (1-2\omega)\,\,\frac{|\rho|^2}{2r^2}\left(\frac{\phi}{\xi}\right)^2
\right\}\, .
\label{betaS}
\eeq
Here we used the condition (\ref{largerho}) --- our semilocal string
solution (\ref{strg}), (\ref{3strq}) (or  (\ref{4strq})) was obtained in
the limit of the large string size, $|\rho| \gg\,m_{\gamma ,\,W}^{-1}$. 

We 
will continue to heavily rely on the  condition 
(\ref{largerho})   in our studies of the effective theory on the 
string world sheet. In the opposite case of $\rho\lsim 1/m_{W}$ the string
reduces to a local non-Abelian string, which is well understood 
\cite{HT1,ABEKY,SYmon,HT2}.

To determine the profile function $\omega (r)$
the functional (\ref{betaS}) must be minimized with respect to
$\omega$. Varying (\ref{betaS}) with respect to
$\omega$ one readily obtains 
\beq
\omega=1-\frac{\phi}{\xi}\,.
\label{omegasol}
\eeq
This solution automatically satisfies the boundary conditions (\ref{bcomega}).

Substituting this solution back into the expression for the
sigma model coupling constant (\ref{betaS}) one gets
\beq
\beta_S= \frac{2\pi}{g_2^2}\,\,m^2_{W}\,\, |\rho|^2\,\,\frac12\,
\ln{\frac{L}{|\rho|}}\,.
\label{betaSres}
\eeq
The integral over $r$ in (\ref{betaS}) is logarithmically divergent in the 
infrared. To regularize this divergence we introduce an infrared cutoff $L$
in (\ref{betaSres}). Since this element is very important,
we pause here to discuss it in more detail. The logarithmic divergence is due to
long-range tails of the semilocal string which fall off as powers of $r$ rather than exponentially. The fact that the $\rho$
zero modes of semilocal strings (CP($N-1$) instantons)  are
logarithmically non-normalizable was noted long ago \cite{Narain,Ward,LeSa}.

The problem is ill-defined unless a physical infrared (IR) regularization is provided.
One possibility is to replace an infinite-length string by
that of a finite length $L$. This will also regulate the spread of the string
in the transverse plane \cite{ayu}. However, at the same time, the problem looses its two-dimensional geometry and becomes essentially three-dimensional.
BPS saturation is also lost. In the logarithmic approximation,
when $\ln |\rho |$ is considered to be a large number,
while non-logarithmic terms are neglected, technically the IR regularization by a
finite length of the string remains a viable option.

A more convenient IR regularization, which maintains the
BPS nature of the solution,  can be provided by a small mass difference
$\Delta m_{AB}\neq 0$, see Sect.~\ref{unma}. In this case
$\ln L/|\rho |$ must be replaced by $\ln (|\rho | m_{AB})^{-1}$.

Now let us switch on the $x_k$-dependence of $\rho$, assuming that the $S^a$ 
moduli
are constant ($x_k$-independent). In this case 
the gauge potentials (\ref{Ak}) vanish.
Substituting (\ref{3strq}) in the action (\ref{model})
we readily obtain
\beq
\frac{2\pi}{g_2^2}\, m^2_{W}\, \int dt\, dz\,\,|\pt_k\rho|^2\,\ln{\frac{L}{|\rho|}}\,.
\label{sizeth}
\eeq
This expression is valid  with logarithmic accuracy, i.e. under the assumption that
the logarithm is large and non-logarithmic terms can be neglected.
We will consistently exploit this approximation throughout the paper.

Now, assembling both parts of the action,
 the orientational  and $\rho$  moduli fields kinetic terms
(\ref{betaSres}) and (\ref{sizeth}), we finally get
an effective low-energy theory on the  world sheet of the semilocal
non-Abelian string. Namely, with the logarithmic accuracy,
\beq
S^{(1+1)}_{N_f=3}=\frac{2\pi}{g_2^2}\,m^2_{W}\int dt\,dz\,\left\{
\frac1{4}\,|\rho|^2\,(\pt_k S^a)^2
+|\pt_k\rho|^2
\right\}\,\ln{\frac{L}{|\rho|}}\,.
\label{3th}
\eeq
The two-dimensional theory (\ref{3th}) contains four real degrees
of freedom: two orientational moduli $S^a$ and a complex field $\rho$
related to the size of the semilocal string.
Note that both
kinetic terms  are proportional to the infrared logarithm. 
We see that the coupling constant of the CP(1) part which describes dynamics
of orientational modes is now determined by $m^2_{W}\, |\rho|^2\, \ln{|\rho|}$.
 The geometry of the target space is ${\cal C}^2\times S_2$, where the radius
of $S_2$ is given by the above-mentioned value and depends on the position in 
the complex plane $\rho$.

Repeating the same procedure with the string solution (\ref{4strq}) in the theory
with four flavors we get practically the same action, 
\beq
S^{(1+1)}_{N_f=4}\, =\, \frac{2\pi}{g_2^2}\, m^2_{W}\, \int dt\,dz\,\left\{
\frac1{4}\,|\rho|^2(\pt_k S^a)^2
+ |\pt_k\rho_i|^2
\right\}\,\ln{\frac{L}{|\rho|}}\,,
\label{4th}
\eeq
with the obvious replacement $|\pt_k\rho|^2 \to  |\pt_k\rho_1|^2
+ |\pt_k\rho_2|^2$.
Now, in addition to   two independent 
fields $S^a$, we have four (real) fields $\rho_i$, $i=1,2$,
while the size of the string is given by (\ref{size}).
Equations~(\ref{3th})  and (\ref{4th}) describe the low-energy limit
of the world sheet theory --- they represent a two-derivative truncation
in the derivative expansion. The  zero-mode
 interaction   contains  higher derivatives too. The 
derivative expansion runs in powers of $|\rho| {\pt_k}$ implying that
the effective sigma models  (\ref{3th})  and (\ref{4th}) are applicable
at scales below the inverse string thickness $1/|\rho|$
which, thus,  plays the role of an  ultraviolet (UV) cutoff for the
world-sheet theories (\ref{3th})  and (\ref{4th}).
This is another reason why $\rho$ has to be regularized in the infrared.

\subsection{Unequal masses}
\label{unma}

To get a deeper insight in  physics of the world-sheet theory 
for semilocal strings let us explicitly 
introduce small mass differences for quark flavors in the bulk theory.
Generally speaking, this will lift all internal moduli, introducing a
shallow potential for the moduli fields in the world-sheet theory. 
We will assume, however, that
\beq
\Delta m_{AB}\ll m_W\,.
\label{smallmass}
\eeq
The smallness of  
$\Delta m_{AB}$ ensures that the effective description in terms of a
two-dimensional sigma models is still valid.

To warm up, let us start from the case of the local non-Abelian string, $N_f=2$,
considered previously.
In this case the
mass difference $\Delta m_{12}$ breaks the global SU(2)$_{C+F}$ symmetry 
down to U(1)
generating a VEV of the adjoint field, see (\ref{avev}). Thus, orientational
moduli are  lifted. The corresponding world-sheet theory \cite{SYmon,HT2},
the  CP(1) model
with twisted mass \cite{Alvarez}, still possesses \ntwo.
The action is
\beq
S^{(1+1)}_{N_f=2}=\beta\int dt\ dz \,\left\{\frac12(\pt_k S^a)^2
+\frac{|\Delta m_{12}|^2}{2}\,(1-S_3^2)\right\}.
\label{mo3}
\eeq
It is clearly  seen that the  moduli fields $S^a$ are no longer massless 
(even at the classical level). The theory (\ref{mo3}) has two vacua $S^a=(0,0,\pm 1)$
which correspond to the (1,0) and (0,1) $Z_2$ strings.

Now we can consider  semilocal non-Abelian strings,  starting from
$N_f=3$. Let us first introduce a mass difference $\Delta m_{13}$, while
keeping $m_1=m_2$. Substituting in the action
 the string solution (\ref{strg}) and (\ref{3strq}),
along with the vacuum values (\ref{avev}) for the adjoint fields, 
 we see that the term in the fourth line of Eq.~(\ref{pot})
gives a nonvanishing contribution for $A=3$. A straightforward calculation yields
\beqn
S^{(1+1)}_{N_f=3} &=& \frac{2\pi}{g_2^2}\,m^2_{W}\int dt\,dz\,\left\{
\frac1{4}\,|\rho|^2\,(\pt_k S^a)^2
+|\pt_k\rho|^2
\right.
\nonumber\\[3mm]
&+& 
|m_1-m_3|^2\, |\rho|^2
\Big\}\,\ln \left(|\Delta m_{13} |\, |\rho|\right)^{-1}\, .
\label{3thm13}
\eeqn
As was expected, the $\rho$ zero modes  are lifted. Another
effect seen in (\ref{3thm13}) is that $\Delta m_{13}^{-1} $ 
does indeed assume the role of the infrared cutoff $L$. 
The reason is quite evident:    
if $\Delta m_{13}\neq 0$, the Higgs branch of the bulk theory degenerates down
to isolated vacua, we no longer have
massless fields in the bulk. Therefore,
at very large $r$, i.e.
$r\gg 1/\Delta m_{13}$,
the profile functions (\ref{lump}) in our solution 
modify to acquire an exponential fall-off 
$\sim \exp{(-|\Delta m_{13}|\,r)}$. This exponential tail cuts off  the logarithmic
$r$ integral resulting in  (\ref{3thm13}).

Strictly speaking, if  $\Delta m_{13}\neq 0$  semilocal strings
cease to exist as exact solutions. The vacuum
of the theory (\ref{3thm13}) is at $\rho=0$ where the string 
under consideration becomes local (and our analytic solution
is inapplicable). We
keep a very small $\Delta m_{13}$ in what follows, much smaller than
any other physical parameter of dimension of  mass, in order to cut off the infrared logarithmic
divergences in the world-sheet theory (cf. Eq.~(\ref{3thm13})).
$\Delta m_{13}$ is kept in the argument of the logarithms, but powers of
$\Delta m_{13}$ will be neglected.
As was mentioned,  $\Delta m_{13}\neq 0$  does not spoil the BPS nature 
of the string.

Now let us take into account  $\Delta m_{12}\neq 0$ assuming
 $\Delta m_{13}\ll \Delta m_{12}$. Much in the same way as in the case of the local
strings \cite{SYmon}, we have to modify our solution (\ref{strg}) and (\ref{3strq})
including in the {\em ansatz} an
expression for the adjoint field. Following \cite{SYmon},
\beq
a^a=\frac{m_1-m_2}{\sqrt{2}}\, \left[\delta^{a3}\,b + S^a \,S^3\,(1-b)\right],
\label{adjans}
\eeq
where $b(r)$ is a new profile function
subject to the boundary conditions
\beq
b(0)=0\,,\qquad b(\infty)=1\,.
\label{bbc}
\eeq
 Next, we substitute the string solution
(\ref{strg}), (\ref{3strq}) and (\ref{adjans}) in  the action. Calculation
goes along the same lines as in \cite{SYmon}, therefore it is appropriate to skip   details.
As in the case of the local strings, the minimization procedure yields
\beq
b(r)=1-\omega (r)=\frac{\phi(r)}{\xi}\,.
\label{b}
\eeq
Finally, we obtain
\beqn
S^{(1+1)}_{N_f=3}
&=&
\frac{2\pi}{g_2^2}\,m^2_{W}\int dt\,
dz\,\left\{
\frac1{4}\,|\rho|^2\,(\pt_k S^a)^2
+|\pt_k\rho|^2
\right.
\nonumber\\[4mm]
&+& \left.
\frac{|\Delta m_{12}|^2}{2}\, |\rho|^2\, \left(1-S_3\right)
\right\}\,\ln \left( \Delta |m_{13}|\, |\rho|\right)^{-1}\,.
\label{3thm}
\eeqn
The orientational moduli are lifted by $\Delta m_{12}\neq 0$.
Classically the vacuum manifold of this theory consists of a single branch, 
\beq
S_3=1\,,\quad \rho\;\mbox{arbitrary}.
\label{vacman3}
\eeq
 From the bulk point of view, the
above vacuum is interpreted as the (1,0) semilocal
$Z_2$ string. What about the (0,1) string? 

As a matter of fact, the $Z_2$ symmetry is explicitly broken in (\ref{3thm}),
in contradistinction with the $N_f=2$ case.
There is no semilocal (0,1) string
under the above choice of parameters. Changing the orientation vector $S^a$ from
$S_3=1$ to $S_3=-1$ implies that the second 
rather than the first quark flavor winds at infinity,
see (\ref{sol}). However since $m_2\neq m_3$ 
this is impossible, and no semilocal string of this type 
develops. Of course we still have the local (0,1) string. It corresponds
to $S_3=-1$ and $\rho=0$. It is not seen in the large $\rho$ approximation.

Expanding the potential in Eq.~(\ref{3thm}) around a point on the 
vacuum manifold we see that two fields, $S_1$ and $S_2$, have masses
$(m_1-m_2)$ while $\rho$ remains massless. The (real) dimension of the branch is two,
\beq
{\rm dim}\,H^{N_f=3}_{(1,0)}=2.
\label{dim3}
\eeq
It is not difficult to generalize this analysis to 
 the case $N_f=4$. We assume $m_{12}\neq 0$ while very small
 mass differences $\Delta m_{13}= \Delta m_{24}$
 are kept only in the argument of the logarithm, for the purpose of the IR
 regularization, $\Delta m_{13}\ll \Delta m_{12}$.
Substituting the string solution
(\ref{strg}), (\ref{4strq}) and (\ref{adjans}) in the action we get
\beqn
S^{(1+1)}_{N_f=4}
&=&
\frac{2\pi}{g_2^2}\,m^2_{W}\int dt\, dz\,\left\{
\frac1{4}\,|\rho|^2\,(\pt_k S^a)^2
+|\pt_k\rho_i|^2
\right.
\nonumber\\[4mm]
&+&
\frac{|m_1-m_2|^2}{2}|\rho_1|^2\left(1-S_3\right)
\nonumber\\[4mm]
&+&\left.
\frac{|m_1-m_2|^2}{2}|\rho_2|^2\left(1+S_3\right)
\right\}\,\ln{\frac{1}{|m_1-m_3||\rho|}}\,.
\label{4thm}
\eeqn
This theory classically has two vacuum branches located at
\beq
S_3=1\,,\quad\rho_2=0\,,\quad \rho_1\;\mbox{arbitrary }\,,
\label{1vacman4}
\eeq
 and 
\beq
S_3=-1\,,\quad\rho_1=0\,,\quad \rho_2\;\mbox{arbitrary }\,.
\label{2vacman4}
\eeq
On the first branch
we obtain four (real) states with mass $(m_1-m_2)$, namely, $S_1$, $S_2$ plus
the complex field $\rho_2$, while $\rho_1$ is massless. This branch has dimension
\beq
{\rm dim}\,H^{N_f=4}_{(1,0)}=2\,.
\label{dim41}
\eeq
It corresponds to the (1,0) semilocal  string.

On the second branch we have four massive states too, with mass
$(m_1-m_2)$, namely $S_1$, $S_2$ plus the complex field
 $\rho_1$, while $\rho_2$ is massless. The dimension of this branch is
\beq
{\rm dim}\,H^{N_f=4}_{(0,1)}=2\,.
\label{dim42}
\eeq
It corresponds to the (0,1) semilocal  string.

Concluding this section we would like to emphasize
again  that the effective theories (\ref{3thm}) and (\ref{4thm}) 
 on the world sheet of the semilocal string were derived in the approximation of
 large but not too large values of $\rho$,
\beq
\frac1{m_W}\ll |\rho|\ll \frac1{|m_1-m_3|}\,.
\label{approx}
\eeq
We commented on  the first inequality 
more than once above. The second inequality ensures that
the infrared logarithm in (\ref{3thm}) and  (\ref{4thm}) is a large parameter.
Please, remember that the
theories (\ref{3thm}) and (\ref{4thm}) are  
derived with the logarithmic accuracy.

\section{Semiclassical limit}
\label{semicre}

The general expressions for $S^{(1+1)}_{N_f=3}$ and $S^{(1+1)}_{N_f=4}$
obtained above can be further simplified in the semiclassical limit, using a number of approximations.
Let us reiterate these approximations:

(i) We will work in the window (\ref{approx}). In this window
$\ln \left(\Delta m_{13}\,|\rho|
\right)^{-1}$ is large so that the logarithmic approximation ---
neglecting nonlogarithmic terms compared to logarithmic --- can be
consistently applied. 

(ii) We will work only with the quadratic terms in the derivative expansion.

(iii) We will keep only the leading terms in the expansion in
$\left( m_{W}\,|\rho|
\right)^{-1} $.

(iv) We will assume that $\Delta m_{12}\gg \Delta m_{13},\, \Delta m_{24}$.
Moreover, the parameters $ \Delta m_{13},\, \Delta m_{24}$
are kept in the arguments of logarithms but are neglected elsewhere.
For brevity we will introduce the  notation
\beq
\Delta m \equiv \Delta m_{12}\,.
\eeq
Needless to say $\Delta m_{12}\ll m_{\gamma,\,W}$ so that all fields in the bulk are very heavy compared
to the masses on the string world sheet.
An extra assumption regarding $\Delta m_{12}$ needed at $N_f=3$ will be specified below.

The subsequent derivations are quite straightforward, albeit somewhat cumbersome and involve a few rescalings/redefinitions. Algebraic manipulations to be 
presented below should not overshadow a simple statement that at the very end we obtain  in the semiclassical limit
the theory of free complex fields, two fields for $N_f=3$ and three fields for $N_f=4$.

First, we  introduce a new variable $z$ replacing the $\rho$ moduli,
\beq
z_i=\rho_i\,\left[2\pi\xi\,\ln {\frac{1}{|m_1-m_3|\, |\rho|}}\right]^{1/2}\,.
\label{z}
\eeq
In terms of these new variables $z_i$, applying the logarithmic approximation 
 we  rewrite the world-sheet theories as
\beqn
&&
\underline{ N_f=3\,:}
\nonumber\\[2mm]
&& S^{(1+1)}_{N_f=3}=\int d^2 x\,\left\{
\frac1{4}\,|z|^2\,(\pt_k S^a)^2
+|\pt_k z|^2+
\frac{|\Delta m_{12}|^2}{2}\, |z|^2\left(1-S_3\right)
\right\};
\nonumber\\[5mm]
&& 
\underline{ N_f=4\,:}
\nonumber\\[2mm]
&&S^{(1+1)}_{N_f=4}=
\int d^2 x\,\left\{
\frac1{4}\,|z|^2\,(\pt_k S^a)^2
+|\pt_k z_i|^2
\right.
\nonumber\\[4mm]
&&\qquad 
\quad \,
+
\left.
\frac{|\Delta m_{12}|^2}{2}\, \left[|z_1|^2\left(1-S_3\right)+|z_2|^2\left(1+S_3\right)
\right]\right\}.
\label{4thz}
\eeqn
Here $|z|^2\equiv \sum_i |z_i|^2$.
In terms of $z$ the window (\ref{approx}) becomes
\beq
\frac{2\pi}{g^2_2}\ll |z|^2\ll \frac{\xi}{|m_1-m_3|^2}\,.
\label{approxz}
\eeq

As we will see later, the effective world-sheet theory for the semilocal string
at $N_f=3$ is asymptotically free and generates its own dynamical scale
$\Lambda_{\sigma}$. In this section we will assume that 
\beq
\Delta m\equiv 
\Delta m_{12}\gg\Lambda_{\sigma}\,.
\label{qclimit}
\eeq
This ensures  the weak coupling regime in the world-sheet theory since under 
(\ref{qclimit})
its coupling constant is frozen at the scale  $\Delta m_{12}$.
The world-sheet theory in the case $N_f=4$
turns out to be conformal so the limitation (\ref{qclimit}) does not apply.

As usual, we begin with $N_f=3$. The O(3) sigma model is known to be 
equivalent to
CP(1) model (for a review see e.g.
\cite{nsvz}). The CP(1) model is a U(1) gauge theory of the complex
charged doublet $n^l$ where $l=1,2$, subject to the condition $|n^l|^2=1$.
The relation between $S^a$ and $n^l$ is as follows:
\beq
S^a=\bar{n}_p\,(\tau^a)^p_l\, n^l\,.
\label{nl}
\eeq
In terms of ${n^l}$ the O(3) sigma-model action (\ref{o3}) takes the form
\beq
S^{(1+1)}_{N_f=2}=2\beta \, \int\, d^2 x \, |\nabla_k \,n^l|^{\,2}\,,
\label{cp}
\eeq
where $\nabla_k=\pt_k-iA_k$. The gauge field $A_k$ enters the action
with no kinetic term and can be eliminated, which would lead us 
back to (\ref{o3}). We will 
trade the fields $S^a$ in (\ref{4thz}) for $n^l$.
It is convenient to parametrize $z$ through its modulus and phase,
\beq
z=|z|\exp{( i\gamma )}\,,\qquad 0\le\gamma< 2\pi\,,
\eeq
and 
introduce new  variables
\beq
\vp^l=|z|\,n^l\,,\quad \eta \equiv\vp^{\,l=1}= |z|\,n^1\,,
\quad \chi \equiv\vp^{\,l=2}= |z|\,n^2 \,.
\label{vp}
\eeq
Then Eq.~(\ref{4thz}) implies
\beqn
S^{(1+1)}_{N_f=3}
=
 \int d^2 x \left\{ |\nabla_k \vp^l|^2 +
|\vp^l|^2\, (\pt_k \gamma)^2 
+
|\Delta m|^2 |\chi|^2\right\}\,.
\label{3cp}
\eeqn
Let us have a
closer look at this theory near its Higgs branch. It is immediately
seen that, since the field $\chi$ is massive,  its quantum
fluctuations are small compared to the value of the massless
$\eta$.
Indeed, while $\chi\sim 1$, at the same time $\eta^2\approx
|\vp^l|^2=|z|^2$ is huge due to (\ref{approxz}). 
This means that we can parametrize $\eta$ as
\beq
\eta =|z|\exp{( i\delta)}
\label{etap}
\eeq
where $\delta$ is a phase,  $0\le\delta < 2\pi$.
It is easy to see that under the circumstances we get 
(to the leading order in $1/|z|$) 
\beq
A_k=\left( \pt_k \,\delta\right).
\label{Ak2d}
\eeq
Substituting this expression back in (\ref{3cp}) we arrive at
\beq
S^{(1+1)}_{N_f=3}= \int\,  d^2 x \, \left\{ |\pt_k z|^2 
+|\pt_k \tilde{\chi}|^2
+|\Delta m|^2 |\tilde{\chi}|^2\right\},
\label{3free}
\eeq
where 
\beq 
\tilde{\chi}=\chi\,\exp{( i\delta )}\,.
\eeq
 The field $\eta$ totally 
disappears! Its role was delegated to other terms.
Its absolute value (equal to $|z|$) 
enters the kinetic term for the complex field $z$, while its phase
$\delta$ is gauged away.

Thus, in the semiclassical limit we managed
to reduce the world-sheet theory at $N_f=3$ to a free theory of one massive complex 
field $\tilde{\chi}$ and one massless complex field $z$. 
This is obviously the bosonic sector of an ${\cal N}=2$
sigma model with the twisted mass for the $\tilde\chi$ field.
The massless
field develops a huge VEV $|z|$ on the two-dimensional Higgs branch 
of the theory. This VEV corresponds to a large size $|\rho|$ of the 
semilocal string. 

It is clear that the flat metric in (\ref{3free}) must
have corrections running in powers of 
$2\pi /(g_2^2|z|)$ which, however, must preserve its K\"ahler nature.
We do not see them in our approximation.

Now let us follow the same road to complete our derivation of the world-sheet
theory in the  $N_f=4$ bulk model. Again our starting point is Eq.~(\ref{4thz}).
Rewriting it as a U(1) gauge theory we get
\beqn
S^{(1+1)}_{N_f=4}&=& \int d^2 x \left\{ |\nabla_k \vp^l|^2 +
|\vp^l|^2\,|\, \pt_k u_i|^2
\right.
\nonumber\\[3mm]
&+& \left.
|\Delta m|^2\,  |u_1|^2\, |\chi|^2
+|\Delta m|^2 \, |u_2|^2|\, \eta|^2\right\},
\label{4cp}
\eeqn
where instead of the angle $\gamma$ we now introduce a complex doublet
$u_i$ via
\beq
z_i=|z|\, u_i\,,\qquad |u_i|^2=1\,,
\eeq
  while $\vp^l$'s are defined as in
(\ref{vp}). As was discussed in Sect.~\ref{etsws},  this theory has two
Higgs branches. Now they are located at $\chi=0$, $u_2=0$ and 
 $\eta =0$, $u_1=0$, respectively.

Consider the first Higgs branch. The field $\chi $ is massive, its
fluctuations are of order one. On the contrary, the field $\eta$ develops
a large VEV, $|\eta|=|z|$. 
Using the parametrization (\ref{etap}) and
eliminating
the gauge field by virtue of Eq.~(\ref{Ak2d}) we get (to the leading order in $1/|z|$)
\beqn
S^{(1+1)}_{N_f=4}&=& \int d^2 x \left\{ |\pt_k z_i|^2 
+|\pt_k \tilde{\chi}|^2
+
|\Delta m|^2 \left( |\tilde{\chi}|^2
+ |z_2|^2\right) \right\}\,.
\label{4free}
\eeqn
 The field $\eta$ again disappears in the same sense as in the $N_f=3$ case.

The model (\ref{4free}) is a free theory of two massive fields 
$\chi$ and $z_2$ and one massless field $z_1$ parameterizing the two-dimensional
Higgs branch. As in the $N_f=3$ case, corrections to the flat metric run in powers of $2\pi /(g_2^2|z|)$.
The second Higgs branch of the theory (\ref{4cp}) has the same
free field description, with the interchange
\beq
\{ \tilde\chi ,\,\, z_2\} \leftrightarrow \{ \tilde\eta ,\,\, z_1\}\,.
\eeq
 
\section{Comparison with the  Hanany--Tong formula}
\label{comparison}

As was mentioned in Sect.~\ref{intro},
non-Abelian semilocal strings were analyzed previously  \cite{HT1,HT2}
within a complementary approach based on $D$ branes.
The advantage of this approach is that it is not
limited to the semiclassical approximation. Its disadvantage
is a rather indirect relation
to field theory. To make contact with field theory
it is highly instructive to compare our  field-theoretic results
with those obtained by Hanany and Tong. They 
conjectured that the effective theory on the world sheet of the non-Abelian
semilocal string is given by the strong coupling limit ($e^2\to\infty$)
of a two-dimensional U(1) gauge theory which  in the case of SU(2)$_{\rm color}$ 
under consideration has the form
\beqn
S &=& \int d^2 x \left\{
 |\nabla_{k} \vp^{l}|^2 +|\tilde{\nabla}_{k} z_i|^2
 +\frac1{4e^2}F^2_{kl} + \frac1{e^2}\,
|\pt_k\sigma|^2
\right.
\nonumber\\[3mm]
&+&\left.
2\left|\sigma-\frac{m_l}{\sqrt{2}}\right|^2 \left|\vp^{l}\right|^2 
+ 2\left|\sigma-\frac{m_{i+2}}{\sqrt{2}}\right|^2\left|z_i\right|^2
+ \frac{e^2}{2} \left(|\vp^{l}|^2-|z_i|^2 -\frac{2\pi}{g_2^2}\right)^2
\right\},
\nonumber\\[3mm]
&& 
l=1,2\,,\qquad i=1,...,\,N_e\,,\qquad \tilde{\nabla}_k=\pt_k+iA_k\,.
\label{t}
\eeqn
With respect to the U(1) gauge field the fields $\vp^{l}$ and $z_i$ have
charges  +1 and $-1$, respectively. If only charge $+1$ fields were present, in the limit 
$e^2\to\infty$ we would get a conventional twisted-mass deformed
CP($N-1$) model.
The charge $-1$ fields $z_i$ convert the target space of the corresponding sigma model into a toric variety.
 In the theory with $N_f=3$ we have
one complex field $z_1\equiv z$, with the negative charge, while in the 
case $N_f=4$ we have two negatively charged complex fields, $z_1$ and $z_2$.

The action  (\ref{t})  is  a  bosonic part of a supersymmetric
U(1) gauge theory with four supercharges,
which corresponds to extended \ntwo supersymmetry in two dimensions. In
particular, the last term in (\ref{t})  is a  $D$ term while the
complex scalar $\sigma$ is an \ntwo superpartner of the photon.
The field contents above fits our expectations since 
the string we work with is 1/2-BPS and, therefore,  preserves
four supercharges on its world sheet (half of  supersymmetry in the
bulk theory). 

There is a rather convincing field-theoretic  argument in favor  of the Hanany--Tong
conjecture. Consider the bulk theory (\ref{model}) with $N_f=2$ at the 
singular point (\ref{avev}) on the
Coulomb branch, and 
take the limit $\xi\to 0$ (the point (\ref{avev}) becomes an isolated vacuum
once we switch on the FI parameter $\xi$).
As was shown in \cite{Dorey},
the BPS spectrum of dyons on the Coulomb branch of the
4D bulk theory  (\ref{model}) identically
coincides with the BPS spectrum in the 2D twisted-mass deformed CP(1) model 
(\ref{mo3}). The reason for this coincidence was revealed 
in \cite{SYmon,HT2} (see also Sect.~9 in \cite{SVZ}). 

Consider a monopole of the SU(2) sector of the bulk theory
at $\xi=0$. This is the 't Hooft--Polyakov monopole \cite{thopo} with
mass given by the classical formula $\Delta m_{12}/g_2^2$.
Quantum corrections to this result are determined by the exact Seiberg--Witten
solution \cite{SW2} of the bulk theory. If we now switch on the FI parameter
$\xi\neq 0$, the quarks condense (see Eq.~(\ref{qvev})) triggering formation of 
flux tubes and confinement of monopoles. 
In fact, the magnetic flux of the SU(2) monopole, $4\pi$,
exactly matches the difference of the magnetic fluxes of two elementary $Z_2$
strings, (1,0) and (0,1), see Eq.~(\ref{sol}). The confined monopole is represented by
a junction of these two $Z_2$ strings. In the 
CP(1) world-sheet theory (\ref{mo3})  the confined monopole is seen 
as a kink interpolating between two vacua ($S_3=\pm 1$) of this theory 
\cite{Tong,SYmon,HT2}. 
Although the 't Hooft--Polyakov monopole on the Coulomb branch
looks very different from the string junction of the theory in the Higgs phase,
amazingly, their masses are the same \cite{SYmon,HT2}. This is due to the fact that
the mass of BPS states (the string junction is a 1/4-BPS state) cannot depend on
$\xi$ because $\xi$ is a nonholomorphic parameter. Since the confined monopole
emerges  as a kink of the world-sheet theory, the Seiberg--Witten
formula for its mass should coincide with the exact result for the kink
mass in two-dimensional \ntwo twisted-mass deformed  CP(1) model 
found in  \cite{Dorey}. Thus, we arrive at
the statement of coincidence of the BPS spectra in both theories.
\footnote{ In fact, Dorey deals \cite{Dorey} with the SU($N$) theory at the root of 
the baryonic Higgs branch defined by the condition $\sum_A m_A=0$. However, 
one can check that the  BPS spectra of massive states in  
these two 4D theories are the same
upon identification of $m_A$ of the  SU($N$) theory 
with $m_A-\frac1N \sum_A m_A$
of the U($N$) theory.
Note that there is no Higgs branch in the vacuum (\ref{qvev}) and (\ref{avev}) 
with $N_f=2$ in the U($N$) bulk theory, and 
all states in the bulk are massive.}

We expect this correspondence to be generalizable to
theories with $N_f>N$. The 2D theory (\ref{t}) was studied in 
\cite{DorHolT} (at generic $N_f>N$) where the  BPS spectrum was shown 
to agree with the spectrum of the  U($N$)
four-dimensional QCD with $N_f$ flavors.

The coupling constant in (\ref{t}) is classically identified with the coupling
$2\pi/g^2_2$ of the bulk theory much in the same way as
in $N_f=2$ case, see (\ref{beta}). 
Moreover,  the one-loop coefficient of the $\beta$ function equals 
$2N-N_f$ for both theories. This leads to identification of their
coupling constants in quantum theory and identification of their 
scales, $\Lambda_{\sigma}=\Lambda$ at $N_f=3$, see (\ref{g2three}) and (\ref{g2four}), and conformality at $N_f=4$.
The coincidence of the BPS spectra makes the theory (\ref{t})
a promising candidate for the effective theory on the world sheet of
non-Abelian semilocal strings.
The $D$ term  in (\ref{t}) determines the Higgs branch of this 
theory,
\beq
|\vp^{l}|^2-|z_i|^2 =\frac{2\pi}{g_2^2}\,.
\label{dterm}
\eeq

Now comes the main point point of our comparison.
We will show that in the limit $$|z|\gg 2\pi/g_2^2\,,$$ see Eq.~(\ref{approxz}),
the metric on this Higgs branch becomes flat, and the Hanany--Tong theory (\ref{t})
reduces to our results quoted in Eqs.~(\ref{3free}) and (\ref{4free})
for $N_f=3$ and $N_f=4$, respectively.

Let us start from the case  $N_f=3$. In Eq.~(\ref{t}) we put $m_1=m_3$ 
and $\Delta m= m_1-m_2$.
The Higgs branch of (\ref{t}) is located at
\beq
\sigma=\frac{m_1}{\sqrt{2}}\, \qquad \vp^{l=2}\equiv \chi =0,
\label{3higgs}
\eeq
while $\vp^{l=1}\equiv\eta$ and $z$ are determined by the condition (\ref{dterm}),
\beq
|\eta|^2 -|z|^2 = \frac{2\pi}{g_2^2}\,.
\label{dtermp}
\eeq
These fields has four real components subject to one constraint (\ref{dtermp}).
This gives $4-1-1=2$ for the dimension of the Higgs branch, where we subtract, 
in addition to the constraint (\ref{dtermp}), one U(1) gauge phase. This coincides with the dimension quoted in Eq.~(\ref{dim3}).

Near this Higgs branch the field $\chi $ is massive, with mass $\Delta m$,
and, hence, it does not develop large fluctuations. Therefore, we neglect 
$\chi \sim 1$ compared to   $|z|\gg 1$, as we did in Sect.~\ref{semicre}. Eliminating
the field $\sigma$ by virtue of  its equation of motion yields
$$\sigma=\frac{m_1}{\sqrt{2}}\,[1+O(1/|z|)]\,.$$ Substituting this in (\ref{t}) we get the mass term for $\chi$ identical to that in Eq.~(\ref{3free}).

Now, let us ignore $2\pi/g_2^2$ in the right-hand side 
of Eq.~(\ref{dterm}), along with  
the contribution of $\chi$ in its left-hand side, which is legitimate
since $|z|\gg 2\pi/g_2^2$. This gives $|\eta|=|z|$.
The mean value  of phases,
$$
\frac{1}{2}\, \left({\rm Arg}\, \eta +{\rm Arg}\,z
\right),
$$
can be gauged away by an appropriate choice of gauge in Eq.~(\ref{t}).
At the same time, the relative phase
$$
{\rm Arg}\, \eta -{\rm Arg}\,z
$$
 can be combined with the modulus
$|z|$ to form a complex field with the flat metric. thus, we arrive at the theory
identical to (\ref{3free}) up to corrections in powers of 
$2\pi /(g_2^2|z|)$.

Next, let us  consider the case $N_f=4$. We put $m_1=m_3$, $m_2=m_4$ and 
$\Delta m=m_1-m_2$. Now  the theory (\ref{t}) has two Highs branches located at
\beq
\sigma=\frac{m_1}{\sqrt{2}}\,,\qquad \chi =z_2=0\,,
\label{4higgs1}
\eeq
and 
\beq
\sigma=\frac{m_2}{\sqrt{2}}\,,\qquad \eta=z_1=0\,,
\label{4higgs2}
\eeq
exactly as  our theory (\ref{4cp}). Consider the first Highs branch. 
Near this Higgs branch the
fields $\chi$ and $z_2$ are massive, with mass $\Delta m$. Ignoring these
fields in comparison with large VEV's of the fields $$|\eta|=|z_1|\gg 2\pi/g_2^2$$
and eliminating the field $\sigma$ and the gauge field $A_k$, we arrive at
the theory (\ref{4free}), to the leading order in the parameter $2\pi /(g_2^2|z|)$.

Summarizing,  we confirm the Hanany--Tong conjecture (\ref{t}) by explicit
field-theoretic calculation of the action of the world sheet-theory of the semilocal
non-Abelian strings in the limit of large string size. 
Our derivation   clearly shows that the limits of applicability of the
derivative expansion are set by $\rho^{-1}$, which can be stabilized by 
the quark mass differences. This feature is not seen in the analysis of Hanany and Tong.
It would be extremely 
interesting to check whether the theory (\ref{t}) correctly reproduces 
corrections to the flat metric in powers of the parameter $2\pi /(g_2^2|z|)$.
General arguments in favor of the theory (\ref{t}) summarized at the beginning
of this section (see \cite{HT1,HT2}) indicate that that's plausible. 
However, even if the matching of the power expansions is demonstrated,
the theory (\ref{t})  definitely cannot be the exact answer for
a low-energy theory on the string world sheet of the semilocal string.
It misses corrections 
\beq
O\left(
\left[ \ln\,
\frac{\sqrt{\xi}}{|z|\, |m_1-m_3|}
\right]^{-1}
\right)
\label{logcor}
\eeq
suppressed by a large infrared logarithm.

As we increase $|\rho|$, corrections in powers  of $2\pi /(g_2^2|z|)$ to the flat metric
of the world-sheet theory become exceedingly smaller. However, if we take 
$|\rho|$ too large, the logarithmic corrections to the metric of the type 
(\ref{logcor}) become important. As we have already mentioned, both types of 
corrections are small inside the window (\ref{approx}).

\section{Quantum regime}
\label{qr}

In this section we will  consider the theory (\ref{t}) at $N_f=3$
in the quantum regime
\beq
\Delta m\ll \Lambda_{\sigma}\,,
\label{quantum}
\eeq
eventually taking the limit of  equal quark masses 
which converts quasimoduli into genuine moduli.
The theory (\ref{t}) is studied in \cite{DorHolT};
here we briefly review some results obtained in this paper and translate them 
in terms of the semilocal strings in four dimensions.
 As we already mentioned, the theory 
(\ref{t}) is asymptotically free with the first (and the only)
coefficient of the $\beta$ function
equal to $2N-N_f=1$. The theory runs towards the strong coupling 
in the infrared and
develop its own scale $\Lambda_{\sigma}=\Lambda$. 
At nonvanishing $\Delta m_{12}$
the orientational zero modes $S^a$ are lifted while the size moduli 
$\rho$ remain massless.
They correspond to motion along the two-dimensional Higgs branch of the theory 
(\ref{vacman3}). 

At $\Delta m_{12}\to 0$ the color-flavor SU(2)$_{C+F}$ symmetry is restored
in the bulk theory. {\sl Classically}, we would expect ``spontaneous symmetry
breaking'' on the string world sheet: we would expect the vector $S^a$ to point
 in some 
particular direction and two orientational modes to become massless Goldstone
modes. This does not happen in two dimensions. Quantum effects restore the
SU(2)$_{C+F}$ symmetry on the world sheet and the orientational moduli
never become massless. In fact, the dimension of the Higgs branch
of the world sheet theory remains two \cite{DorHolT} at
$\Delta m_{12}\to 0$. Orientational moduli $n^{l}$ acquire mass of order of
$\Lambda$ much in the same way as  in  CP(1) model. This means that,
although the size $\rho$ of 
the semilocal string can have arbitrary values, the orientational vector $S^a$
does not have any particular direction. It smeared all over. The string
is in a highly quantum  non-Abelian regime at $\Delta m_{12}\to 0$. 
This is in one-to-one correspondence with the case 
of local non-Abelian strings  \cite{ABEKY,SYmon} occurring in the  in 
the theory with $N_f=N$.

The quantum regime in the theory (\ref{t}) with $N_f=4$ is quite different.
It is conformal, no dynamical scale develops. The quasiclassical analysis
of Sect.~\ref{etsws} can be extended to include the limit $\Delta m_{12}\to 0$,
provided the coupling constant $g_2$ is small. In particular, we see that
the Higgs branch of the theory gets enhanced in the limit $\Delta m_{12}\to 0$.
In fact, two two-dimensional Higgs branches (\ref{1vacman4}) and 
(\ref{2vacman4}) fuse to become a connected six-dimensional vacuum manifold,
\beq
{\rm dim}\,H^{N_f=4}=6\,.
\label{dim4}
\eeq
It corresponds to all four size moduli fields plus  two orientational moduli 
fields becoming  massless. This indicates that the string is in the ``classical
non-Abelian  regime,'' namely,  the orientation vector $S^a$  points in some
particular  direction. The SU(2)$_{C+F}$ group is not restored on the world 
sheet by quantum effects.
This regime does not occurs for local
non-Abelian strings in quantum theory \cite{ABEKY,SYmon}.

In conclusion, we stress
that  in both cases, $N_f=3$ and $N_f=4$, the  size zero moduli $\rho_i$ of 
the  semilocal non-Abelian string are {\sl not} lifted by interactions with the
orientational moduli  in the quantum regime. This means that 
taking account of quantum effects does not change the fact that the size of the 
semilocal  non-Abelian string is arbitrary provided all mass differences
are switched off. As was discuss in Sect.~\ref{ssac} this effectively leads 
to deconfinement.

Let us note, that IR-conformal theories, such as the one in Eq.~(\ref{t}) with $N_f=4$
and finite $e^2$,
were studied in \cite{W97,OogVa}. If $m_l =0$ at $l=1,2,3,4$, and in the limit
$2\pi/g^2_2\to 0$,
these theories were shown to develop both the Higgs and Coulomb branches
with distinct values of the 
Virasoro central charge.
Moreover, a tube metric for the field $\sigma$ was shown to be generated
at one loop, upon integrating out the matter fields. It is interpreted as 
a long tube connecting the Higgs and Coulomb branches.

\section{Conclusions}
\label{conclu}

In this paper we considered a benchmark bulk theory in four-dimensions:
 \ntwo supersymmetric QCD with the gauge group U($N$) and
$N_f$ flavors of fundamental matter hypermultiplets (quarks). 
The nature of the BPS strings in this benchmark theory crucially depends
on $N_f$. If $N_f\geq N$  and all quark masses are equal,
it supports non-Abelian BPS strings which have internal (orientational)
moduli associated with
rotations of the color magnetic  flux in the  non-Abelian group SU($N$). 
If $N_f>N$ these strings become semilocal, developing additional 
moduli related to (unlimited) variations  of their
transverse size. 

Using the U(2) gauge group
with $N_f=3,4$ flavors as an example, we derive an
effective low-energy theory on
the (two-dimensional) string world sheet. 
Our derivation is field-theoretic; it is direct and explicit in the sense that
we first analyze the Bogomol'nyi equations for string-geometry solitons,
suggest an {\em ansatz} and solve it at large $\rho$.
Then we use this solution to  obtained the world-sheet theory.

Our result considered in the semiclassical limit confirms
the conjecture made previously by Hanany and Tong that this theory is 
\ntwo supersymmetric U(1) gauge theory  in two dimensions with $N$
positively and $N_e$ negatively charged matter multiplets and the
Fayet-Iliopoulos term determined by the four-dimensional coupling constant.
We discuss physics of this model and conclude that its Higgs branch is
not lifted by quantum effects. This means that the width of the string
can freely grow. As a result, such strings cannot confine.

Our analysis of infrared effects shows that, in fact, the derivative expansion
can make sense only provided the theory under consideration 
is regularized, e.g. by the quark mass differences. The world-sheet action discussed in this paper
becomes a {\em bona fide} low-energy effective action
only if $\Delta m_{AB}\neq 0$.

\section*{Acknowledgments}

We are grateful to Arkady Vainshtein for very useful discussions.

The work of M.S.  was 
supported in part by DOE grant DE-FG02-94ER408.
The work of A.Y. was  supported 
by  FTPI, University of Minnesota, and by Russian State Grant for 
Scientific Schools RSGSS-11242003.2.

\end{document}